\documentclass{jaa}
\usepackage{natbib}
\usepackage{graphicx}

\begin{document}\sloppy

\title{A Comparison  of the UV and HI Properties of the Extended UV (XUV) Disk Galaxies NGC 2541, NGC 5832 and ESO406-042}


\author{M.~Das\textsuperscript{1}, Yadav, J.\textsuperscript{1}, Patra, N.\textsuperscript{2}, Dwarakanath, K.S.\textsuperscript{2}, McGaugh, S.S.\textsuperscript{3}, Schombert, J. \textsuperscript{4}, Rahna, P.T.\textsuperscript{5}, and Murthy, J.\textsuperscript{1}}
\affilOne{\textsuperscript{1}Indian Institute of Astrophysics, II Block, Koramangala, Bangalore 560034, India.\\}
\affilTwo{\textsuperscript{2}Raman Research Institute, Sadashivanagar, Bangalore 560034, India.\\}
\affilThree{\textsuperscript{3}Case Western Reserve University, Cleveland, OH 44106, USA\\}
\affilFour{\textsuperscript{4}University of Oregon, Eugene, OR 97403, USA\\}
\affilFive{\textsuperscript{5}CAS Key Laboratory for Research in Galaxies and Cosmology, Shanghai Astronomical Observatory, Shanghai, 200030, China.\\}


\twocolumn[{

\maketitle

\corres{mousumi@iiap.res.in}

\msinfo{...}{to appear in JAA 2021}

\begin{abstract}
We present a UV study of 3 extended UV (XUV) galaxies that we have observed with the UVIT and the GMRT. XUV galaxies show filamentary or diffuse star formation well beyond their optical disks, in regions where the disk surface density lies below the threshold for star formation. GALEX observations found that surprisingly 30\% of all the nearby spiral galaxies have XUV disks. The XUV galaxies can be broadly classified as type~1 and type~2 XUV disks. The type~1 XUV disks have star formation that is linked to that in their main disk, and the UV emission appears as extended, filamentary spiral arms. The UV luminosity is associated with compact star forming regions along the extended spiral arms. The star formation is probably driven by slow gas accretion from nearby galaxies or the intergalactic medium (IGM). But the type~2  XUV disks have star formation associated with an outer low luminosity stellar disk that is often truncated near the optical radius of the galaxy. The nature of the stellar disks in type~2 XUV disks are similar to that of the diffuse stellar disks of low surface brightness galaxies. The star formation in type~2 XUV disks is thought to be due to rapid gas accretion or gas infall from nearby high velocity clouds (HVCs), interacting galaxies or the IGM. In this paper we investigate the star formation properties of the XUV regions of two type~2 galaxies and one mixed XUV type galaxy and compare them with the neutral hydrogen (HI) emisison in their disks. We present preliminary results of our UVIT (FUV and NUV) observations of NGC 2541, NGC 5832 and ESO406-042, as well as GMRT observations of their HI emission. We describe the UV emission morphology, estimate the star formation rates and compare it with the HI distribution in these type~2 and mixed XUV galaxies.
\end{abstract}


}]


\doinum{12.3456/s78910-011-012-3}
\artcitid{\#\#\#\#}
\volnum{000}
\year{0000}
\pgrange{1--}
\setcounter{page}{1}
\lp{1}

\section{Introduction}

Star formation is one of the main processes driving galaxy evolution. It can be triggered externally by interactions or mergers with nearby galaxies, by gas infall from companion galaxies or by gas accretion from the intergalactic medium (IGM). Gas infall cools galaxy disks leading to disk instabilities and star formation. Star formation is also internally driven by secular evolution, this includes global disk instabilities such as spiral arms and bars \citep{kataria.das.2018}. Most of the star formation activity in spirals is confined to the inner optical disk ($<$5 kpc), where the stellar and gas surface densitites are large enough for instabilities to set in \citep{kennicutt.1989}. Earlier studies indicate that the disk surface brightness is exponential and has a sharp disk truncation radius, where the stellar mass surface density falls and the star formation rate declines \citep{vanderkruit.searle.1981} . However, deep H$\alpha$ studies have shown that there is star formation in the extreme outer disks of some nearby galaxies, such as NGC 628 and NGC 6946 \citep{ferguson.etal.1998}, where H$\alpha$ emission is detected beyond the R$_{25}$ optical disk radii.  

The most definitive evidence, however, of outer disk star formation came from the ultraviolet (UV) observations of nearby galaxies by GALEX \citep{martin.etal.2005,morrissey.etal.2007}. The GALEX surveys showed that $\sim$30\% of all spiral galaxies at distances of $<$40 Mpc have UV emission from HII regions in their outer disks \citep{gildepaz.2007}. This is surprisng as in these regions the stellar density declines. Such galaxy disks are called extended UV (XUV) disk galaxies \citep{thilker.etal.2007}. Good examples are M83 and NGC 2090 that show extended spiral features in their extreme outer, low density disks \citep{thilker.etal.2005}. Their existence supports models of inside out star formation, where star formation occurs in the extreme (optically) low surface brightness (LSB) zones of galaxy disks resulting in disk building and galaxy growth. XUV disks are also one of the best examples of star formation in low density environemnets \citep{krumholtz.mckee.2008} and the stellar disk initial mass function (IMF) in such environments is known to be deficient in massive stars \citep{bruzzese.etal.2020}. Hence, XUV disks are interesting to study -- both for understanding disk growth as well as for studying star formation in underdense environments. 

The XUV disk galaxies have been classified earlier in the GALEX surveys into 3 types, type~1, type~2, and one mixed category \citep{thilker.etal.2007}. The type~1 XUV disks have UV bright complexes arising from spiral arms, arm segments or complexes in the faint, outer parts of stellar disks. Hence they can be associated with the disk spiral structure and are thought to arise from slow gas accretion in the outer disk regions \citep{deblok.etal.2014}. They span over all Hubble types of spiral galaxies. The type~2 XUV galaxies have a blue color (FUV-NIR) in the extreme outer parts of their galaxy disks. These regions appear as optically low in surface brightness and similar to the disks of LSB galaxies \citep{honey.etal.2016,das.2013}. The LSB zone lies outside most of the K band luminosity of the disk ($\sim$80\%). Type~2 XUV galaxies are mainly gas rich, late type spiral galaxies that have low stellar masses. Their XUV disks are thought to arise from rapid gas accretion and the ensuing star formation. This is unlike Type I XUV disks where the gas accretion is thought to be slow. There are many XUV disks that show both type~1 and 2 properties, they are called mixed XUV disks. About 20\% of the spirals in the GALEX atlas of nearby galaxies have type I XUV disks and 10\% have type II XUV disks.

A common property of all XUV galaxies and of those that show outer disk star formation is that they are rich in neutral hydrogen gas (HI) and their HI disks are usually more extended than normal galaxies \citep{bigiel.M83.2010}. They are overall nearly twice as gas rich compared to non-XUV galaxies, but on the other hand there are many non-XUV galaxies that have extended gas disks but do not support XUV star formation (e.g. NGC 2915) \citep{thilker.etal.2007}. So being gas rich in the outer disk is a necessary but not sufficient condition for XUV star formation. Some XUV disks are interacting with nearby companions, but not all. In such cases the HI is disturbed and there maybe HI tidal tails or bridges \citep{kaczmarek.etal.2012}. In this study we focus on the star formation in type~2 XUV galaxies, and try to understand the nature of star formation in their extended LSB disks \citep{mcgaugh.etal.1995}. We use both UVIT observations and HI observations to compare the distribution of star forming complexes (SFCs) within the gas distribution. The long term goal is to understand what drives XUV star formation and why only some extended HI disks show XUV star formation. In the following sections we describe our observations, our results and then discuss the implications of our study for understanding star formation in the outer disks of galaxies.

\begin{table*}[htb]
\tabularfont
\caption{List of Galaxies and Parameters}\label{tableExample} 
\begin{tabular}{lccccccc}
\topline 
Galaxy      & RA, Dec                     & Class    & v$_{sys}$        & Diameter             & Distance  & Scale                    & XUV  \\
            & (J2000)                     &          & (km/s)           & ($^{\prime\prime}$)  &  (Mpc)    &  (kpc/$^{\prime\prime}$) & type \\
\hline
ESO406-042  & 23 02 14.2, -37 05 01.4 & SAB(s)   & 1365             & 138.0                & 18.6      & 0.090                    & type~2 \\
NGC 2541    & 08 14 40.1, +49 03 42.2 & SA(s)cd  &  548             & 396.4          & 11.9      & 0.058                    & type~2    \\
NGC 5832    & 14 57 45.7, +71 40 56.4 & SB(rs)b? &  447             & 222.9                &  9.5      & 0.046                    & mixed type \\                 
\hline
\end{tabular}
\end{table*}

\begin{figure*}
\centering
\includegraphics[height=.28\textheight]{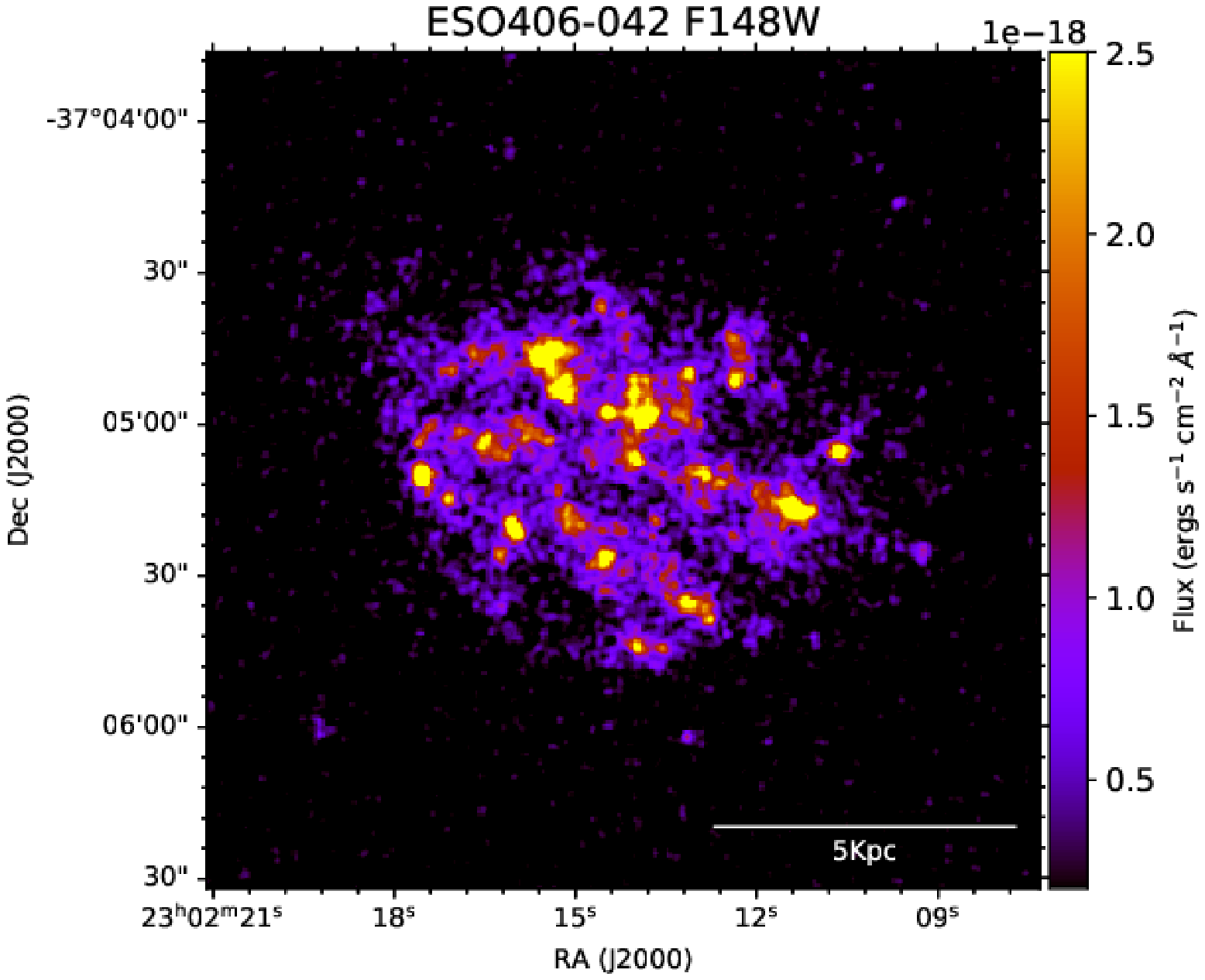}
\includegraphics[height=.28\textheight]{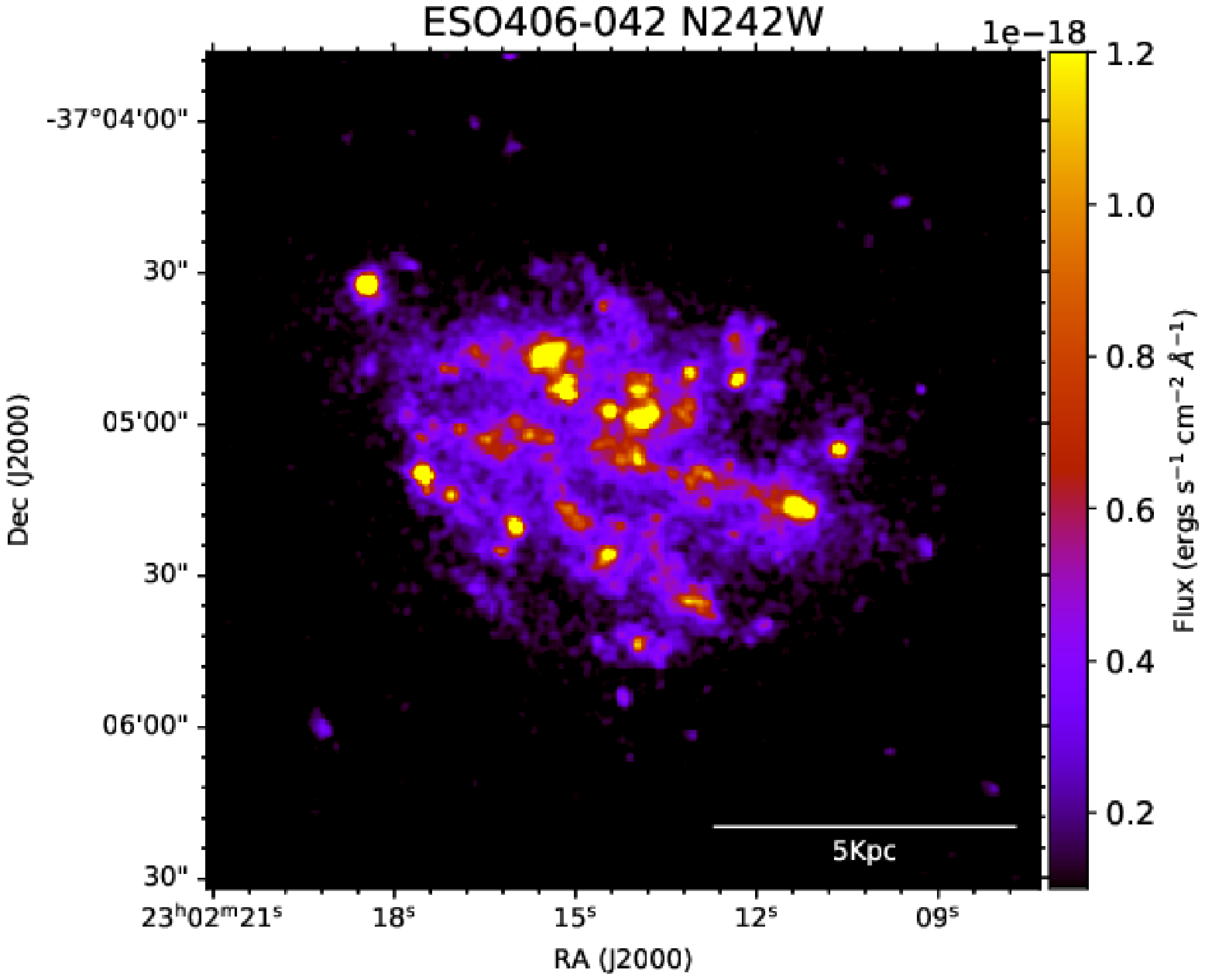}
\includegraphics[height=.28\textheight]{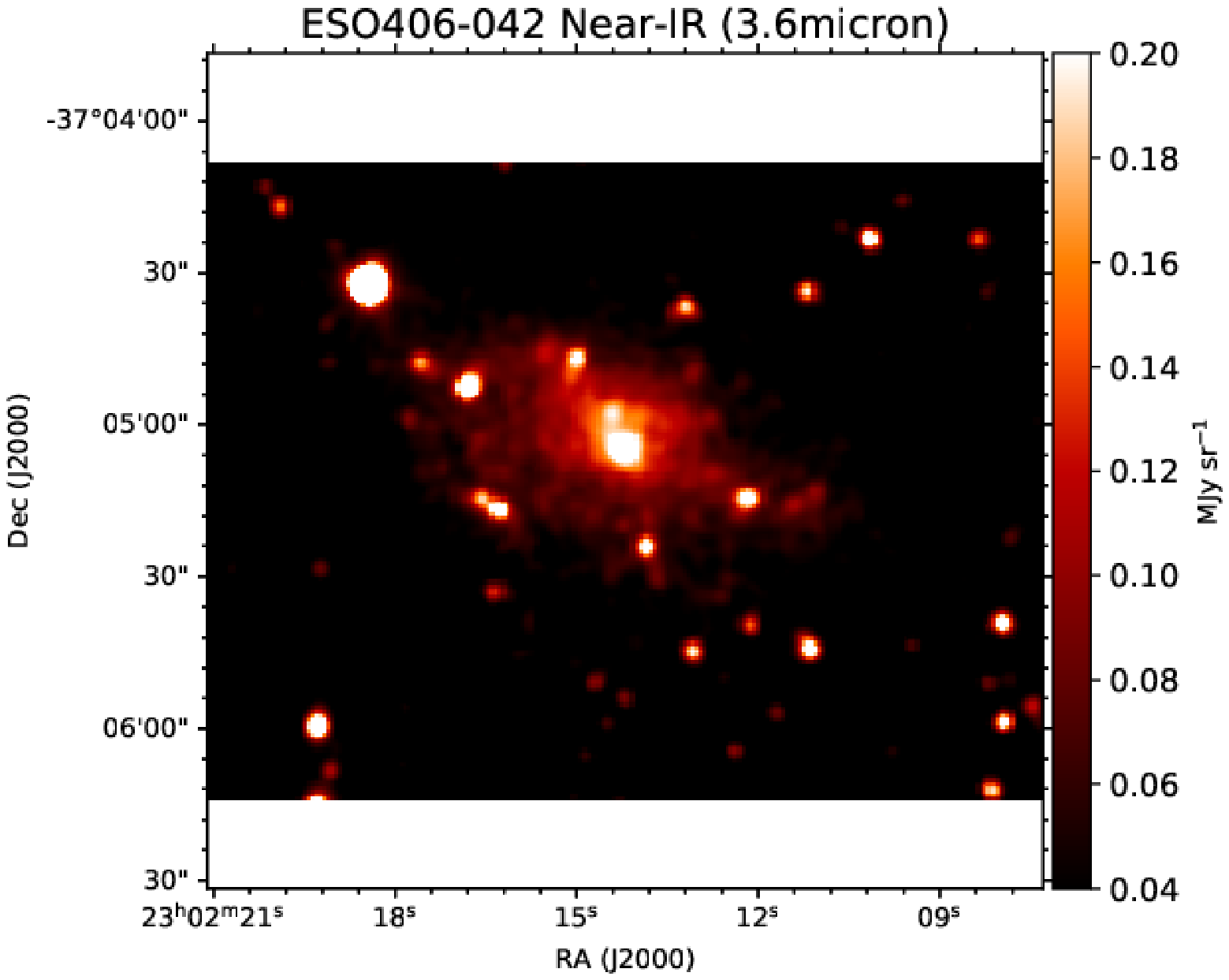}
\includegraphics[height=.28\textheight]{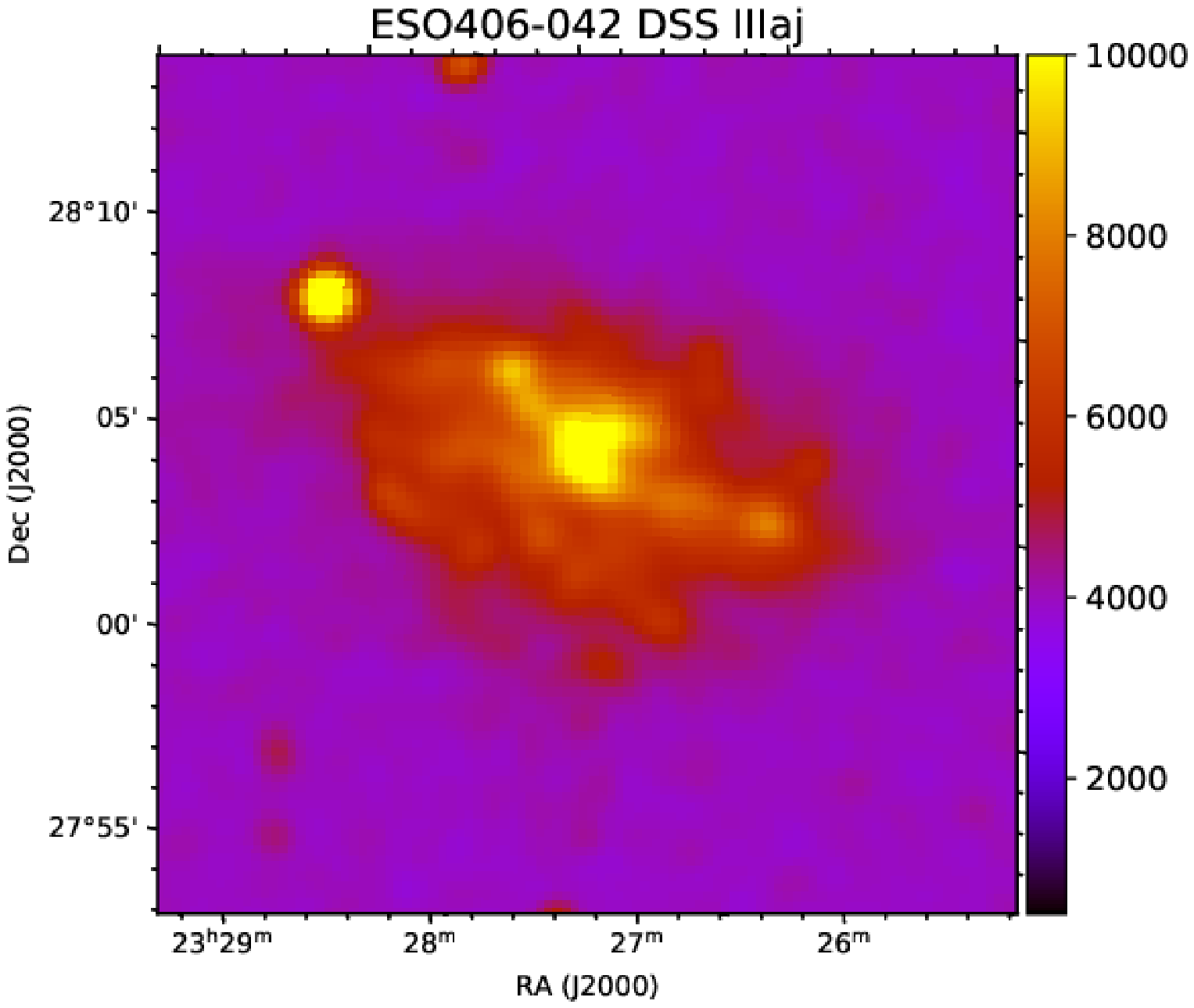}
\caption{The above panel shows the UV and optical images of ESO406-042. From top left clockwise (i)~the UVIT FUV image, (ii)~the UVIT NUV image, (iii)the DSS IIIaj optical image which is based on the photographic data obtained using the U.K. Schmidt telescope and (iv)~the Spitzer 3.6$\mu$m image.  }
\end{figure*}

\begin{figure*}
\centering
\includegraphics[height=.28\textheight]{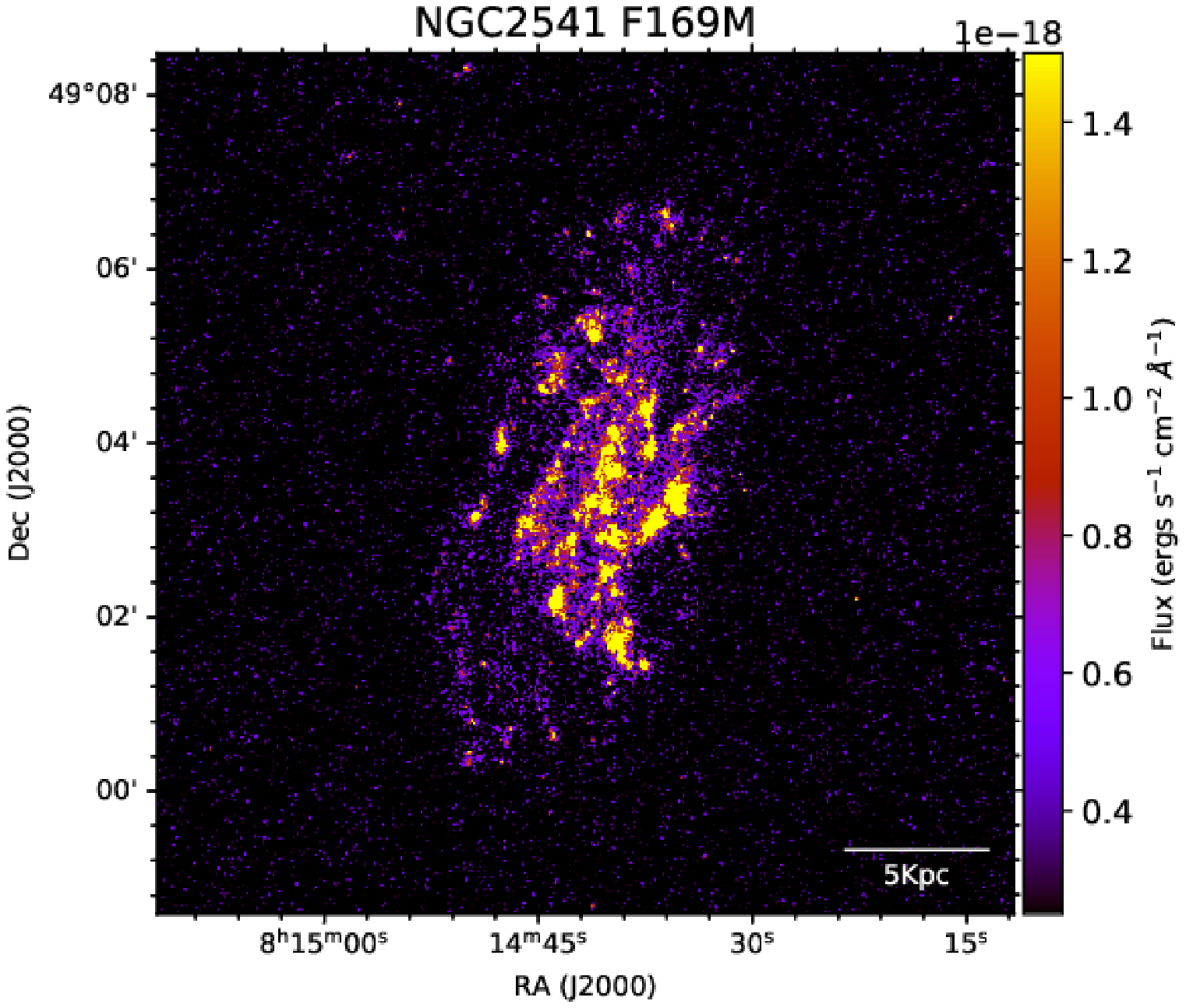}
\includegraphics[height=.28\textheight]{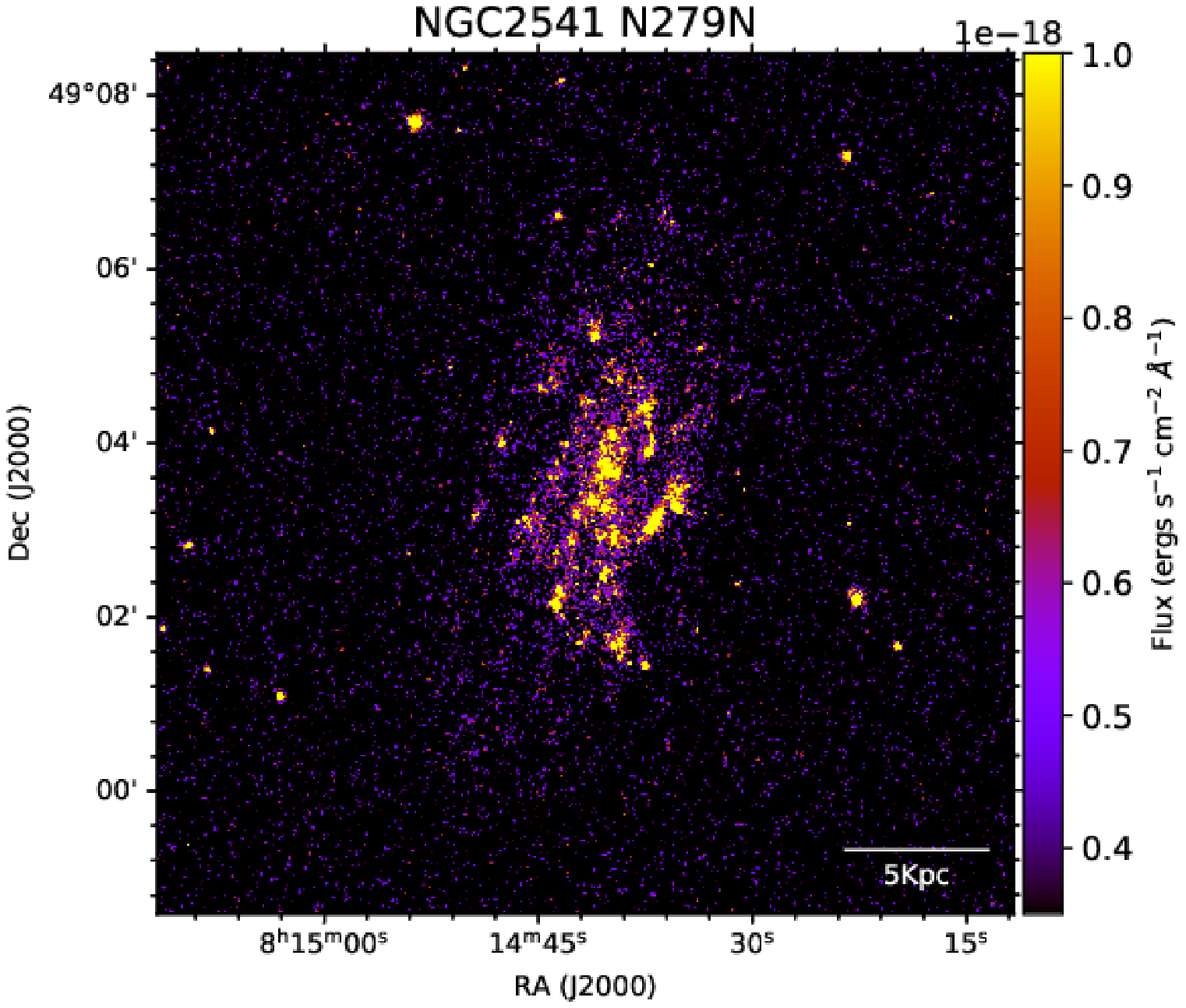}
\includegraphics[height=.28\textheight]{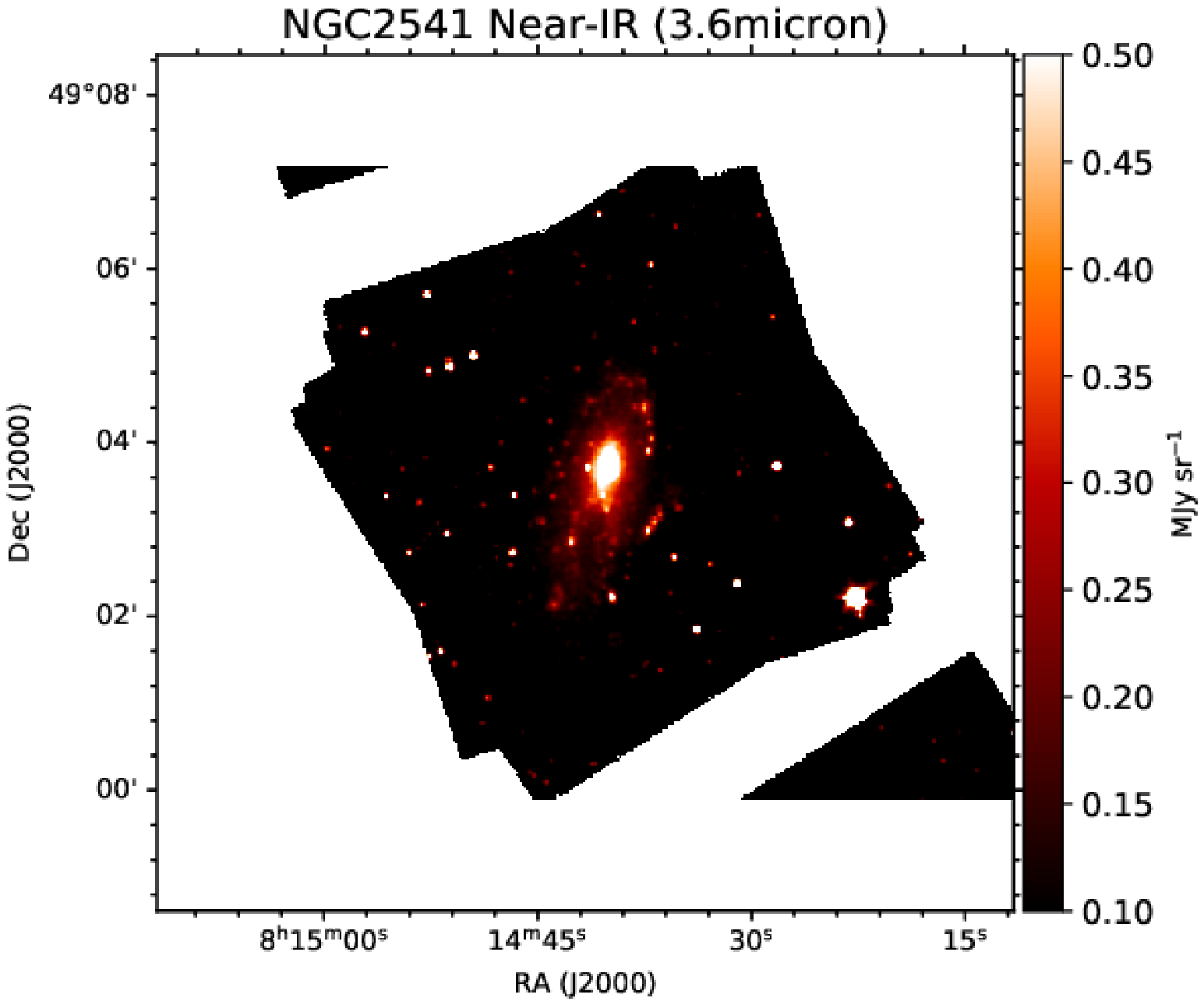}
\includegraphics[height=.28\textheight]{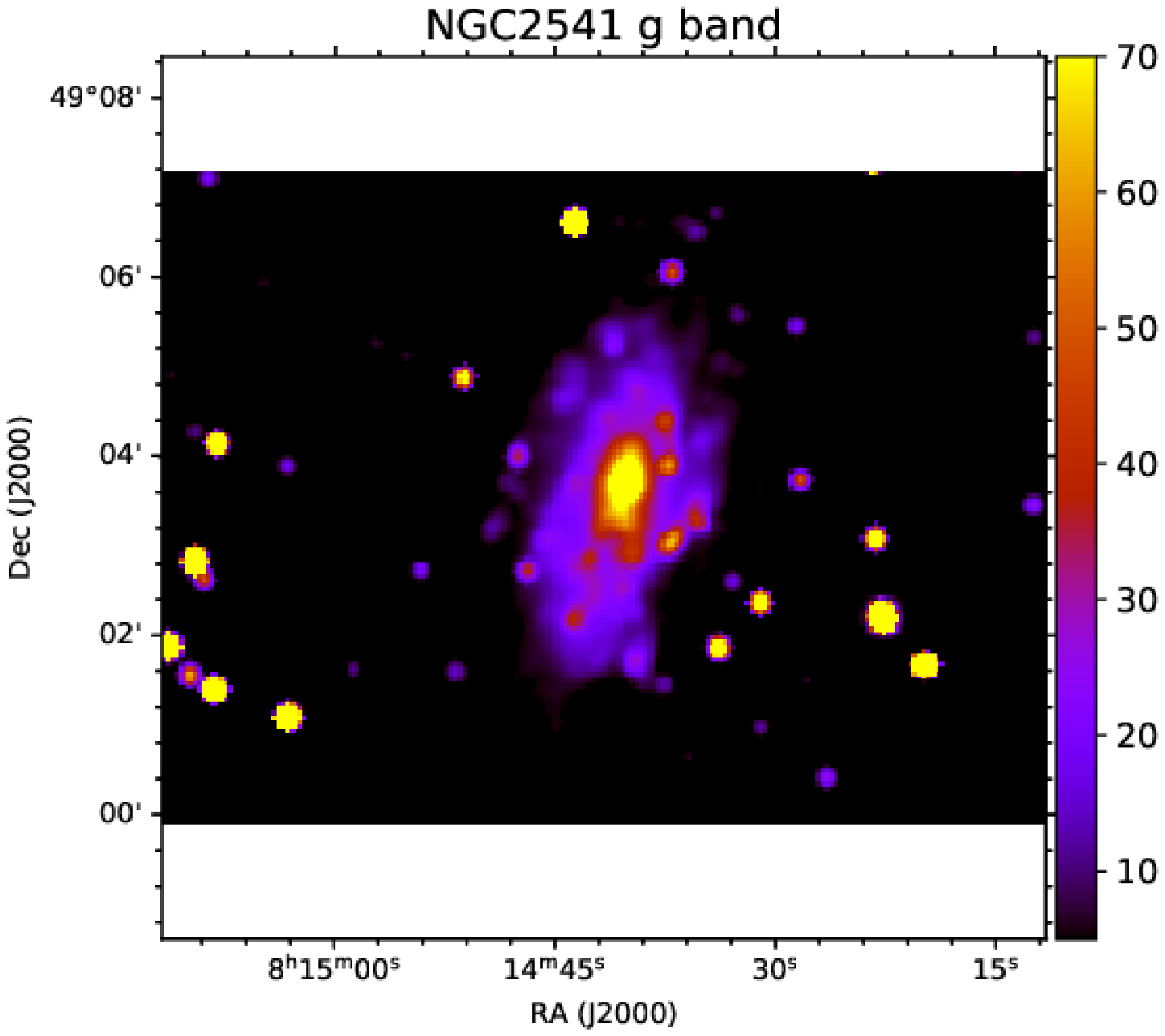}
\caption{The above panel shows the UV and optical images of NGC 2541. From top left clockwise (i)~the UVIT FUV image (ii)~the UVIT NUV image, (iii)~the SDSS g band optical image and (iv)~the Spitzer 3.6$\mu$m image.  }
\end{figure*}

\begin{figure*}
\centering
\includegraphics[height=.28\textheight]{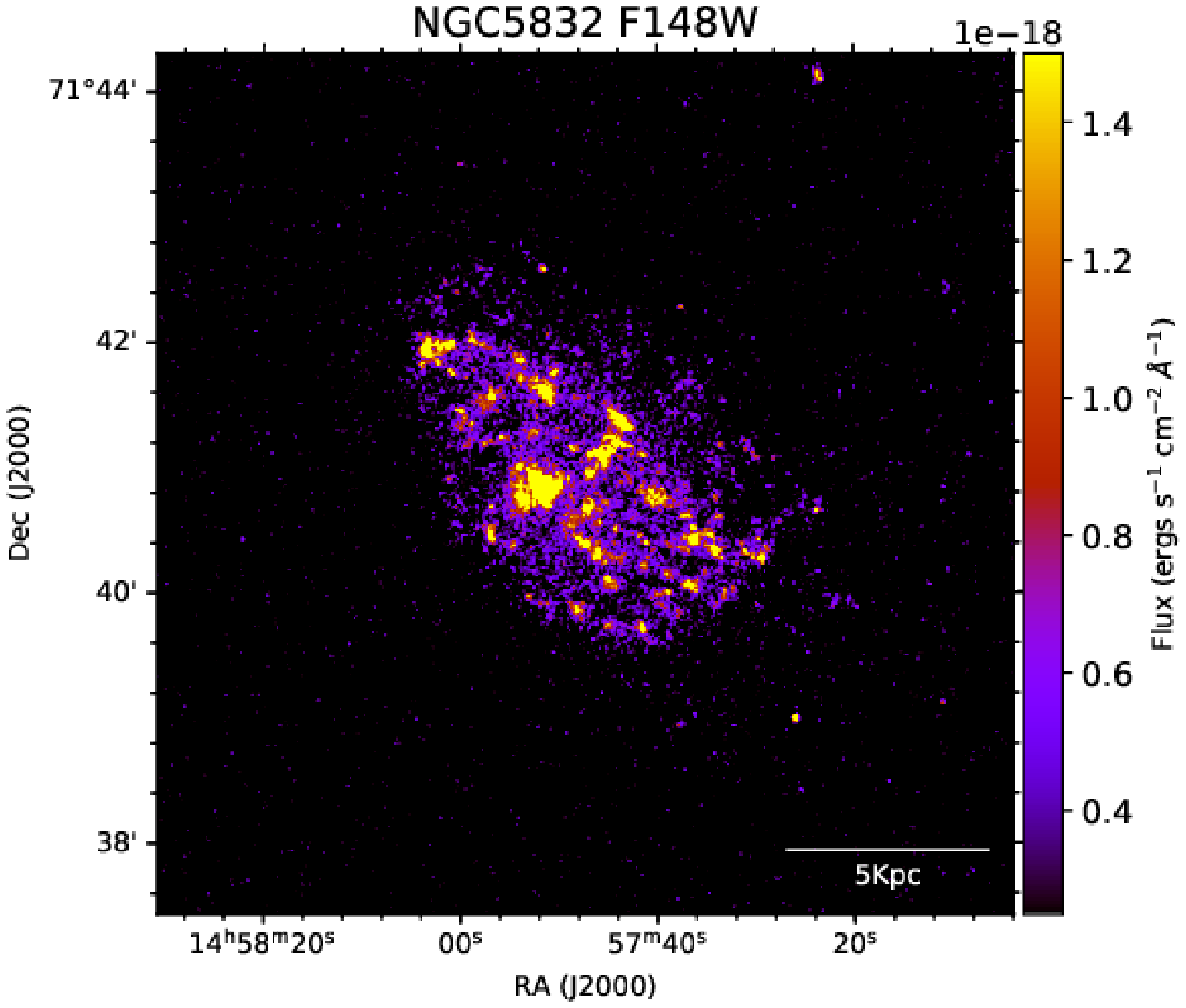}
\includegraphics[height=.28\textheight]{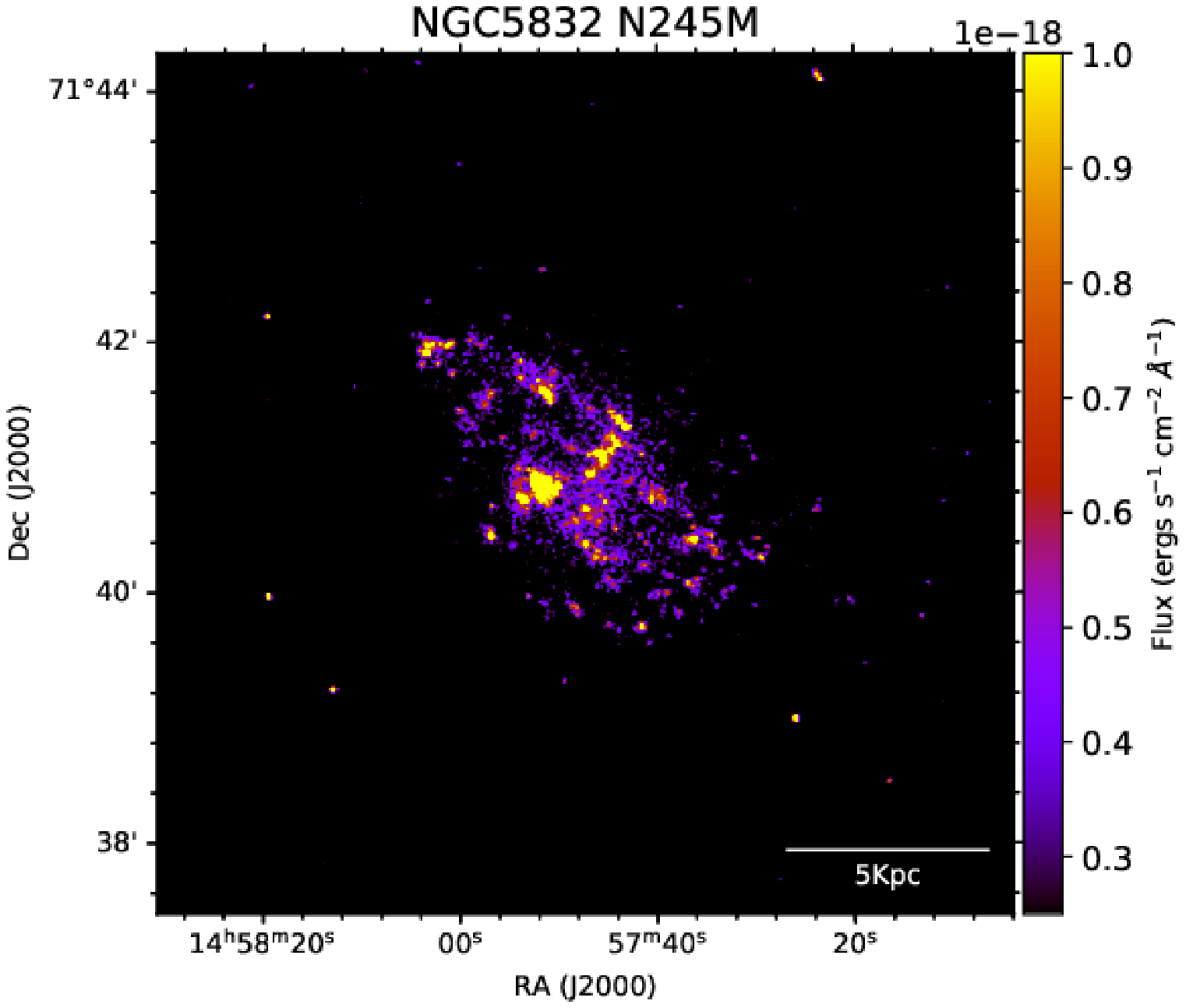}
\includegraphics[height=.28\textheight]{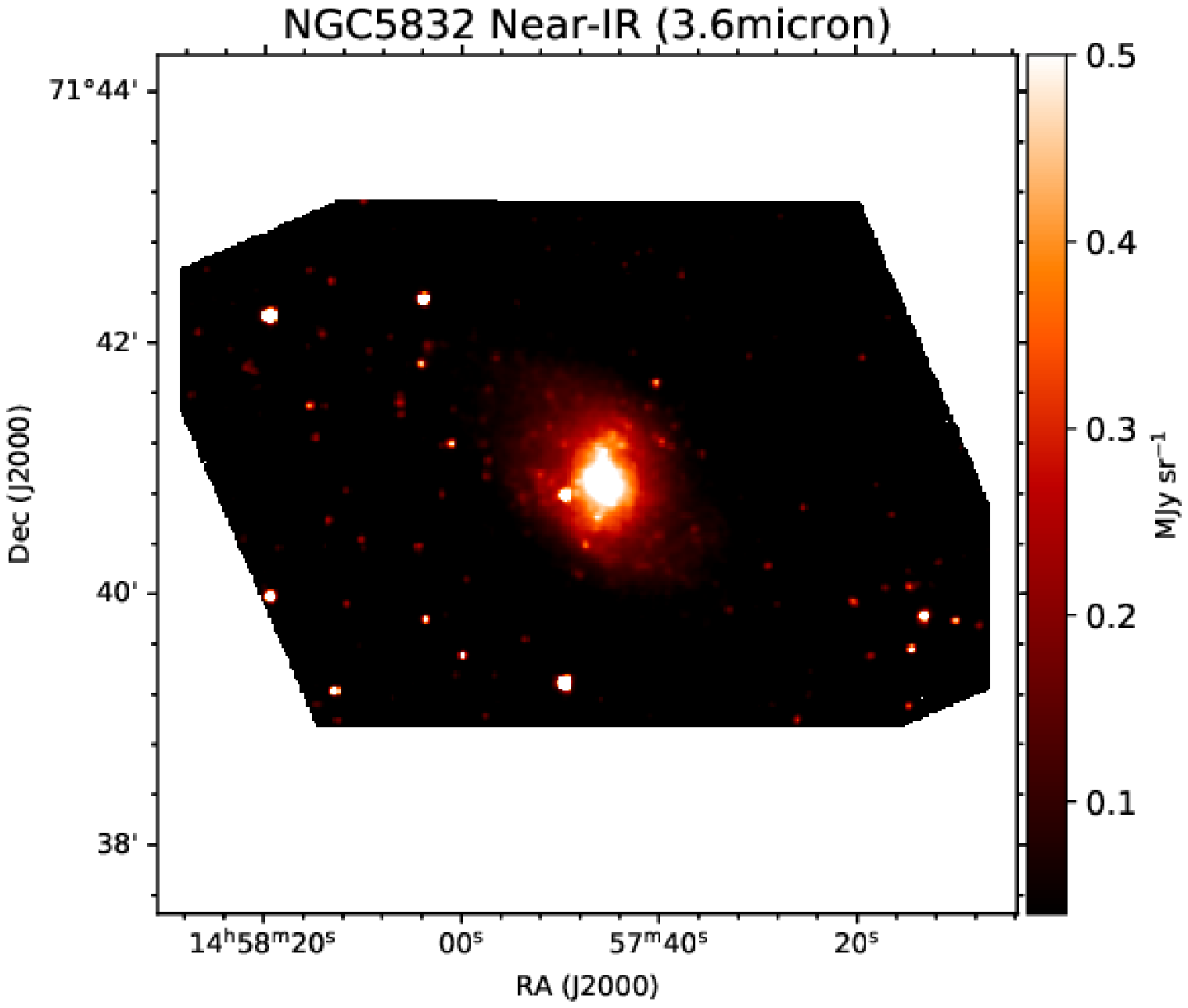}
\includegraphics[height=.28\textheight]{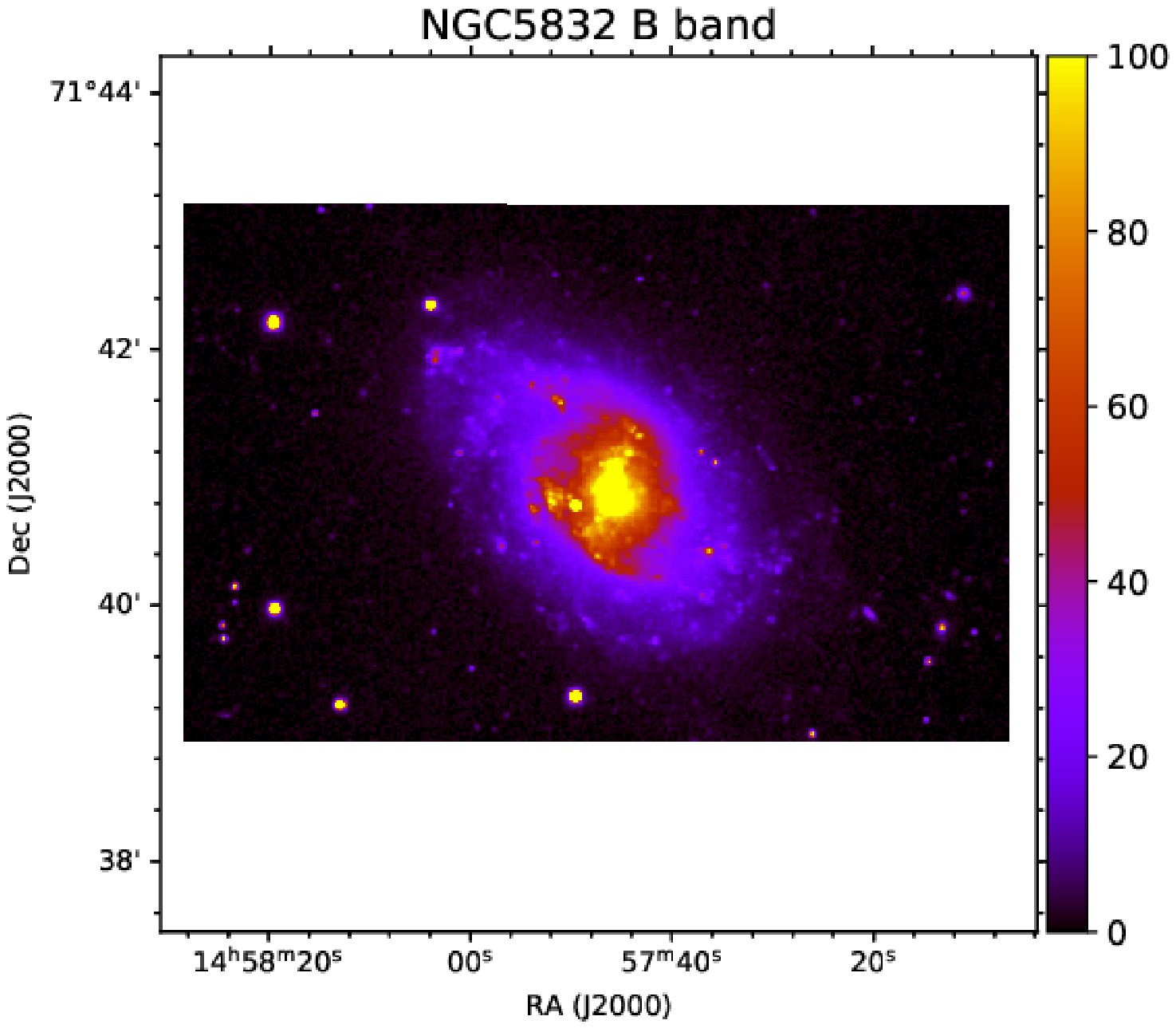}
\caption{The above panel shows the UV and optical images of NGC 2541. From top left clockwise (i)~the UVIT FUV image (ii)~the UVIT NUV image, (iii)~the B band optical image from the Spitzer Legacy Optical Photometry Survey \citep{cook.etal.2014}, and (iv)~the Spitzer 3.6$\mu$m image.}
\end{figure*}

\section{Galaxy Sample}
All our targets have been observed by GALEX and form part of the surveys of XUV disk galaxies \citep{lemonias.etal.2011}. Two targets are type~2 XUV galaxies and one is a mixed type XUV galaxy. We have focussed largely on the type~2 XUV disks as we want to understand the rapid gas accretion onto the outer disks of galaxies and see its relation to HI morphology. The galaxies are bright, nearby sources which means that the high spatial resolution of UVIT can resolve star forming regions in their XUV disks \citep{rahna.etal.2018}. Below we discuss the general properties of the galaxies, which are also listed in table~1.\\
ESO406-042 : This is a relatively small galaxy with a stellar disk of size 12.6~kpc (Table~1). The Spitzer 3.6$\mu$m image shows that there is a small but bright bulge in the galaxy center, but the stellar disk appears relatively diffuse and hence maybe low in surface density (Figure~1). It is thus not surprising that the galaxy is classified as a type~2 XUV galaxy, since the disk or outer XUV region is LSB in nature. The optical image shows that there are star forming knots distributed over the entire disk but there is no clear spiral structure. Two small arms can be seen in the optical image, but the overall appearance suggests a flocculant spiral structure rather than a strong one. \\
NGC 2541 : This is the largest XUV galaxy in our sample and it has a disk diameter of 21.19~kpc (Table~1). It has a bright oval shaped bulge and an extended LSB disk, which is clearly detected in the B band image (Figure~2). The stellar disk in the 3.6$\mu$m image does not show any clear spiral structure, but faint spiral arms can be seen in the B band image. The outer disk shows surprisingly clear spiral arms in UV but they appear more flocculent in nature, and do not appear to connected with the inner disk which is typical of the type~1 XUV galaxies.\\  
NGC 5832 : This is also a relatively small galaxy with a disk diameter of 10.25~kpc (table~1). It is classified as a barred galaxy but the bar is difficult to trace in the Spitzer 3.6$\mu$m image (Figure~3) and it is relatively small compared to the disk size. The inner stellar disk is bright at 3.6$\mu$m but is very diffuse in the outer parts. This is similar to disks in type~2 XUV galaxies. The outer disk also appears to have a different positions angle with repect to the inner one, this maybe due to the effect of the bar. There appears to be a faint spiral structure which is clearest in the B band image. 

\begin{table*}[htb]
\tabularfont
\caption{Summary of UVIT observations and the UV Fluxes }\label{tableExample} 
\begin{tabular}{lcccccc}
\topline
Galaxy      &   Band  & Filter   & Exposure time  & PSF ($^{\prime\prime}$) &  Cycle,      & Flux \\
            &         &          &   (Ks)         & (image resolution)      &   date       & (ergcm$^{-2}$s$^{-1}$A$^{-1}$)\\
\hline
ESO406-042  & FUV     & CaF2     & 1.994              & 1.1                     & A04-053, 14 Oct, 2017   & 2.58*10$^{-14}$\\
     "      & NUV     & Silica   & 2.010              & 1.0                     &    "                    & 1.15*10$^{-14}$\\
NGC 2541    & FUV     & sapphire & 1.992              & 1.0                     &  A02-075, 21 Dec, 2016  & 1.32*10$^{-13}$\\
     "      & NUV     &  N2      & 2.008              & 1.1                     &    "                    & 6.56*10$^{-14}$\\
NGC 5832    & FUV     &  CaF2    & 1.972              & 1.1                     & A04-053, 03 Mar, 2018   & 6.56*10$^{-14}$\\
     "      & NUV     &  NUVB13  & 1.691              & 1.0                     &    "                    & 2.64*10$^{-14}$\\                 
\hline
\end{tabular}
\tablenotes{The point spread function or PSF of the image is measured by fitting a gaussian to a background bright star and taking the FWHM. For our UV images this is an approximate measure of the resolution of the image. }
\end{table*}

\section{Observations and Data Analysis}
We performed FUV and NUV imaging observations of the galaxies listed in Table~1 using the UVIT on board the AstroSat Satellite \citep{Kumar.etal.2012}. The details of the filters and observational time are listed in Table~2. The instrument has two co-aligned Ritchey Chretien (RC) telescopes, one for FUV (1300-1800 A), and another for both the NUV (2000-3000 A) and visible bands. The UVIT is capable of simultaneously observing in all three bands, and the visible channel is used for drift correction. The UVIT has multiple photometric filters in both UV bands and the field of view of around 28$^{\prime}$ and a spatial resolution$\leq$1.5$^{\prime\prime}$ which is much better than GALEX \citep{rahna.etal.2017}. In our images the spatial resolution is $\geq 1^{\prime\prime}$ and the image resolution is determined from measuring the PSF of a background star (see Table~2 for our image values). A comparison of our UVIT and GALEX observing times are given in Table~4. Our UVIT FUV and NUV observing times are $\sim$2ks whereas the GALEX exposure times are of the order of 0.1 to 0.3Ks. Hence our UVIT observations of these 3 XUV galaxies are far deeper than GALEX. 

We downloaded the UVIT level 1 data of the 3 galaxies from the Indian Space Science Data Centre (ISSDC). We used CCDLAB \citep{postma.leahy.2017} to reduce the level 1 data. The CCDLAB has a graphical user interface to reduce the UVIT data and corrects for field distortions, and also does drift corrections. We did the astrometry on the UVIT images using GAIA data. We used a tool in CCDLAB which can match GAIA sources with UVIT sources and do the astrometry. We determined the background counts and then subtracted it from the image. We calculated the counts per second (CPS) from a galaxy by summing over the CPS from a circular aperture of radius equal to the optical radius (R$_{25}$) of the galaxy (see Table~2). To calculate the flux we used the formula : 
\begin{equation}
Flux (erg s^{-1}cm^{-2}A^{-1}) = CPS*(UC)
\end{equation}
\noindent
where UC is the unit conversion factor and is derived from the equation $ZP=[-2.5 log(UC\times\bar\lambda^2) - 2.407]$ \citep{tandon.etal.2017}, and $ZP$ is the zero point of the filter \citep{tandon.etal.2020}. To calculate the Flux in Jansky (Jy) which is listed in Table~4, we converrted the CPS to AB magnitude using the formula, $m(AB) = -2.5 log10(CPS) + ZP$. After calculating the magnitude we corrected for the Milky way extiction for each filter and then converted the Milky Way extinction corrected magnitudes to fluxes in Jansky using the formula, Flux(Jy)=$10^{-({M(AB)}_{corr} - 8.9)/2.5)}$, where $M(AB)_{corr}$ is the corrected absolute AB magnitude. The star formation rate (SFR) was calculated using the FUV luminosities in Jy, using the formula from \citet{salim.etal.2007} and the filter information from Tandon et al. (2020). For the galaxies ESO406-042 and NGC5832 this was straight forward because the FUV fluxes were obtained using the CaF2 filter which has a wavelength range similar to GALEX. Hence the formula from Salim et al., which is applicable to GALEX observations, could be used. However, for NGC 2541 the filter used was sapphire. Hence, we calibrated the CPS in this filter with the GALEX observations using 7 stars in the field of the galaxy. It should be noted that since the UV fluxes have only been corrected for Milky Way extinction and not for the dust extinction arising from within the individual galaxies, i.e without including the IR contribution, our SFR estimates are possibly lower than the true values.

As part of this project, we observed all the three galaxies in our sample with the Giant Meter wave  Radio Telescope (GMRT) during November-December, 2018 \citep{patra.etal.2019} (see Table~3 for details). All these galaxies were observed with the GMRT Software Backend (GSB) with either a 16.67 or 4.16 MHz bandwidth divided into 512 channels. This results in a spectral resolution of $\sim$ 32/8 kHz ($\rm \sim 6.8/1.7~km~s^{-1}$); $\approx$ 8 hours of telescope time was used for every observation. This led to an on-source time of $\sim$5 hours per source. Standard flux calibrator on the GMRT sky (3C48, 3C147, or 3C286) was observed for $\sim 10-15$ minutes in the beginning and at the end of every observing run. Secondary calibrators near the target source (within 10$^{\circ}$) were observed for $\sim$6 minutes at $\sim$30 mins cadence during the observations. The details of the observations are presented in Table 3.

The data were analyzed using the classic Astronomical Image Processing Software (AIPS). For every observing run, the data were first inspected for dead antennas, and visibilities coming from these antennas were removed. Next, the visibilities were further inspected and edited to exclude any bad data due to Radio Frequency Interference (RFI), sudden gain variation, antenna malfunction, etc. Flux and phase calibration was done using the AIPS task {\tt CALIB} taking the primary and secondary calibrators as reference. A comparison of the recovered flux and the secondary calibrator's reported flux in the NVSS catalog indicates consistency within $\sim 10-15\%$. After the flux and the phase calibrations were done, we performed a bandpass calibration to account for the gain variation in frequency. To do that, we used secondary calibrators, as they were observed with a higher cadence. We do not attempt to perform any self-calibration as the number of bright sources in the field is limited in L-band (1.4 GHz). We then applied all these calibrations to the target sources (our galaxies) and separated their data using the task {\tt SPLIT}. This data contains signals from both the continuum and the HI line. However, as we are only interested in HI~emission from these galaxies, we removed the continuum before we imaged the line emission. To do that, we first averaged all the visibilities from the line-free channels in a visibility cube. This average visibility data was then imaged to produce continuum maps of our galaxies. These maps only contain emission from the continuum sources in the galaxies. We used these maps to produce visibility cubes only for the line. We used the AIPS task {\tt UVSUB} to subtract the continuum from the parent visibility cubes. However, in the presence of a strong continuum (e.g., for ESO 406-042), {\tt UVSUB} may not be able to remove it completely. In such cases, we further used the AIPS task {\tt UVLIN} to fit the residual continuum as a first-order polynomial and subtracted it. We thus produced a visibility cube without having any continuum. The frequency axes of these visibility cubes were then shifted to a heliocentric velocity frame using the AIPS task {\tt CVEL}. After the frequency standardization, all the continuum free visibility cubes were imaged channel by channel using the AIPS task {\tt IMAGR}. To maximize the S/N and not resolve the diffused HI~emission, we use a UV tapering of 5 kilo-lambda and a natural weighting scheme during imaging. For our observations, we find a typical single channel RMS of $\sim 3$ mJy/Beam. The image cubes are then further used to produce moment maps using the AIPS task {\tt MOMNT}. We note that for NGC 5832, the data were heavily corrupted by RFI, so no usable spectral cube could be made. Hence, here we only present the results from NGC 2841 and ESO 406-042. In Figure~4, we show the global HI spectra (blue dashed lines) for our sample galaxies, as obtained by our GMRT observations. For comparison, we also plot the same as obtained by single-dish observations (solid red lines).
 
\begin{table*}[htb]
\tabularfont
\caption{GMRT observation details}\label{tableExample} 
\begin{tabular}{lcccccccc}
\topline
Galaxy      & Date of      & On-source   & Observing  & Number of  & Spectral      & Flux & Phase  & RMS/chan  \\
            & Observation  &  time (Hr)  &  BW (MHz)  & channels   &  width (km/s) & Cal  &  Cal   &  (mJy/bm)   \\
\hline
ESO406-042  & November 25, 2018    & 4.5                  &  16.66             & 512                & 6.9                   & 3C48     & 0010-418  & 3.0 \\
NGC 2541    & December 23, 2018    & 5.5                  &  4.16              & 512                & 1.8                   & 3C147    & 0713+438  & 3.2  \\
NGC 5832    & December 23, 2018    & 5.5                  &  4.16              & 512                & 1.8                   & 3C286    & 1400+621  & $---$  \\
\hline
\end{tabular}
\end{table*}

\begin{figure*}
\centering
\includegraphics[height=.22\textheight]{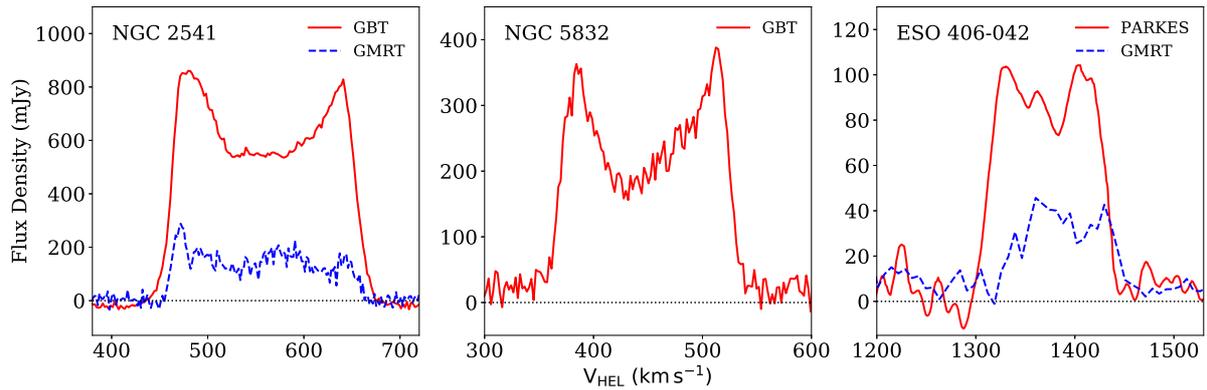}
\caption{The above panel shows the HI line emission for the 3 galaxies. They have been derived from our GMRT interferometric observations and also compared with single dish fluxes from other telescopes. }
\end{figure*}

\begin{figure*}
\centering\includegraphics[height=.23\textheight]{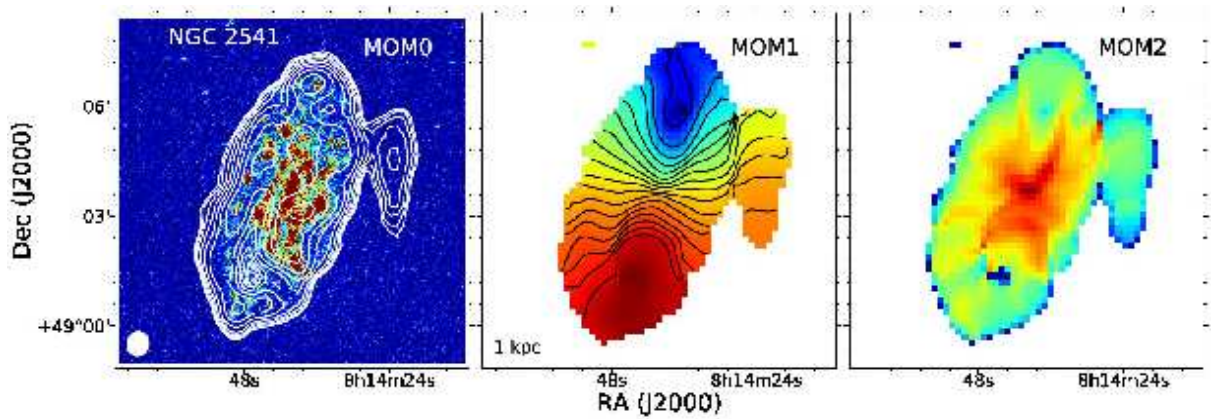}
\caption{The panel on the left shows the HI emission for the galaxy NGC 2541 observed with the GMRT. The plot on the left is the HI intensity image or moment0 map. The contours are (1,1.4,2,2.8,....) x 8 x 10$^{19}$ atoms/cm$^2$, i.e., the contours start from 8$\times$10$^{19}$ and successive contours are separated by sqrt(2). The one in the middle is the HI velocity field or the moment1 map with contours from 470 km/s to 640 km/s. Contour separation is by 10 km/s. The figure on the right is the moment2 map of NGC 2541 with Minimum = 0 km/s, Maximum = 22 km/s
Note that the HI emission has an arm like structure west of the galaxy center. This appears as an extended spiral structure in the deeper Westerbrock telescope images.}
\end{figure*}

\begin{figure*}
\centering
\includegraphics[height=.24\textheight]{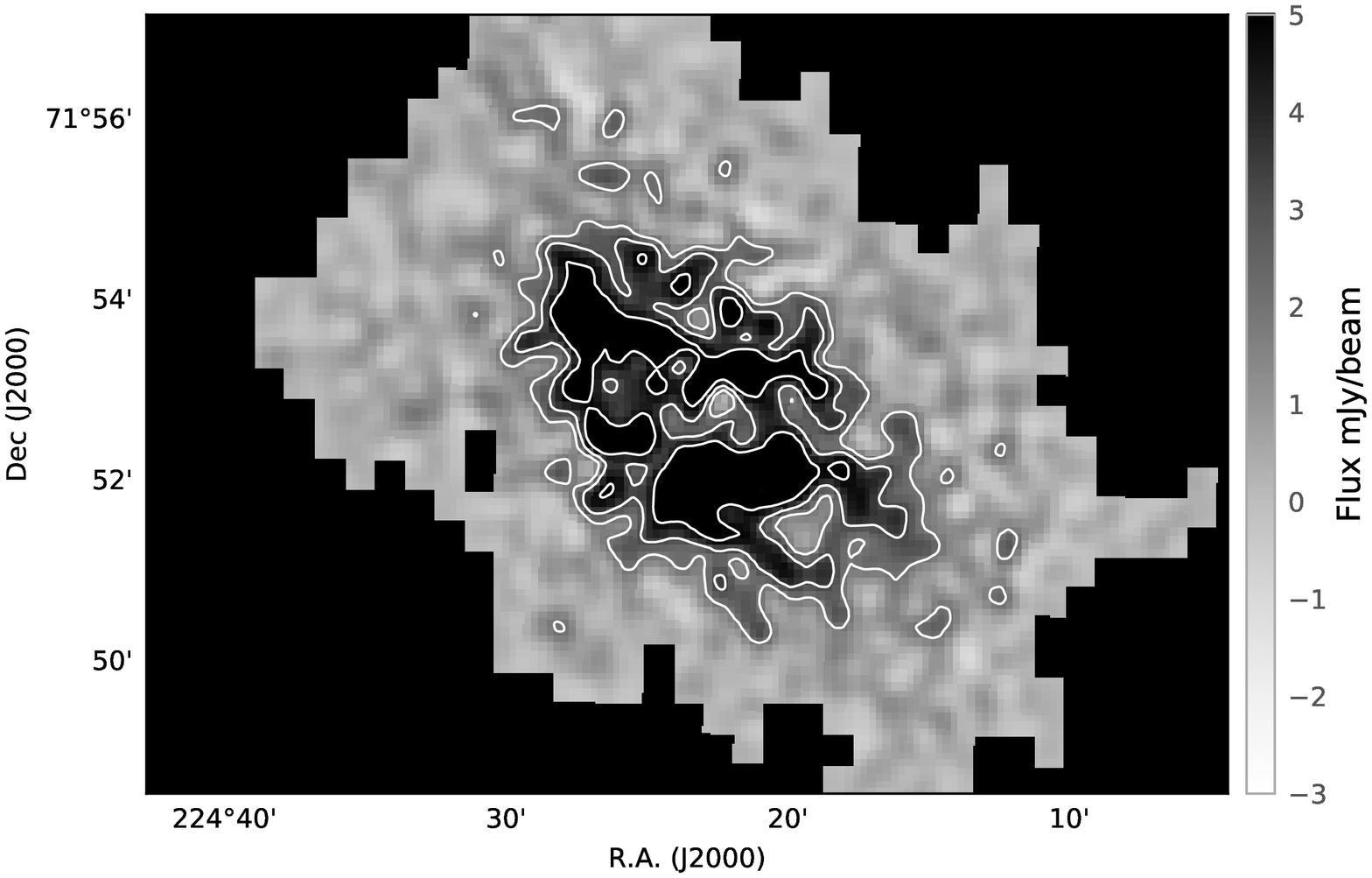}
\includegraphics[height=.24\textheight]{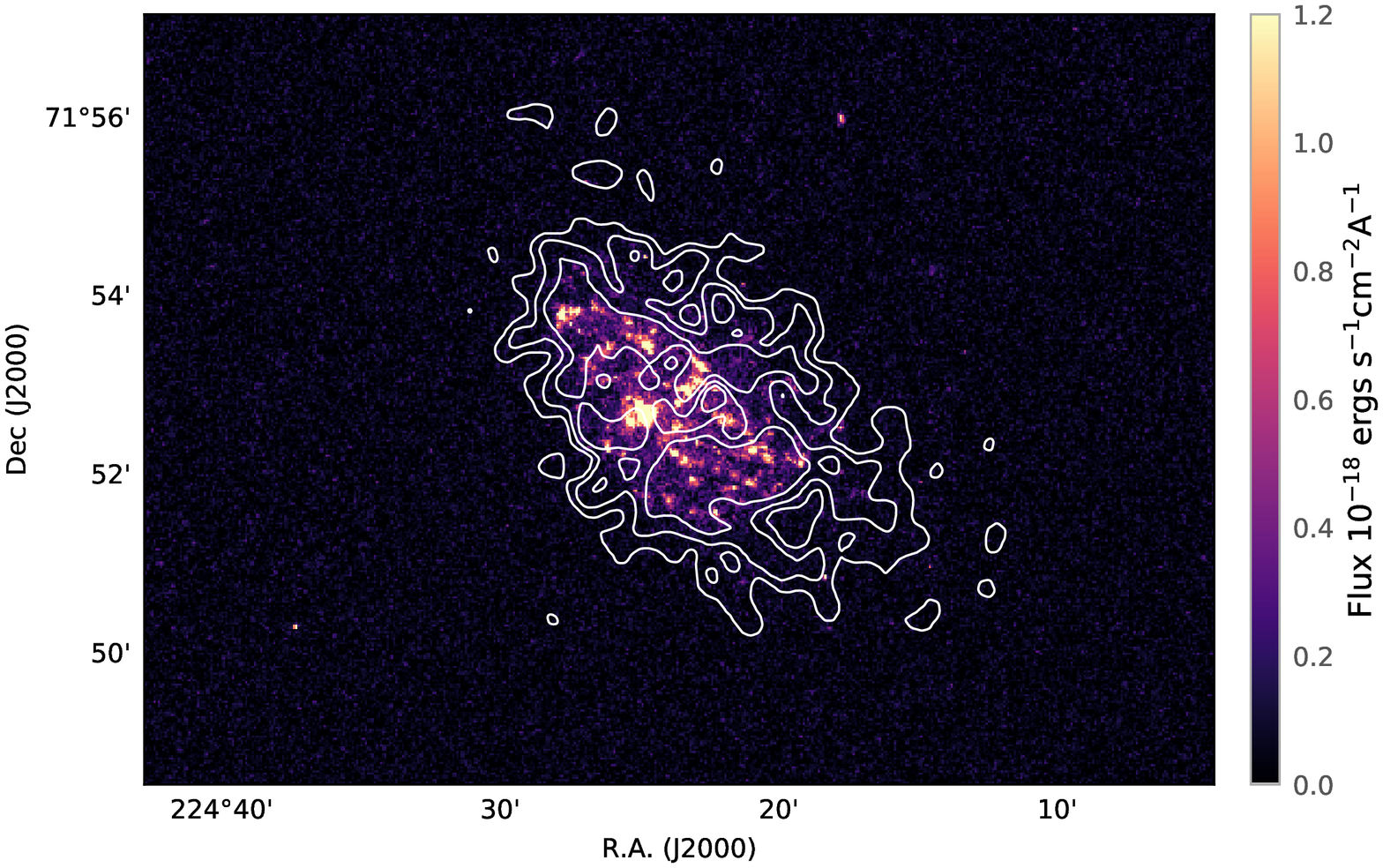}
\caption{The figure on the left shows the HI emission for the galaxy NGC 5832 observed with the Westerbrock Synthesis Radio Telescope (WSRT) \citep{vanderhulst.etal.2001}. The plot on the right is the FUV image of NGC 5832 with contours of HI emission overlaid. The contours are 1/5, 1/3 and 1/2 of the peak intensity which is 9.7~mJy/beam and the beam is $10.8^{\prime\prime}\times 10.1^{\prime\prime}$. The contours follow the FUV emission quite well but also extend slightly beyond the emisison. }
\end{figure*}

\begin{table*}[htb]
\tabularfont
\caption{Comparison of Fluxes between UVIT and Galex, and the Star formation rates (SFR) derived from UVIT}\label{tableExample} 
\begin{tabular}{lccccccc}
\topline
Galaxy      &   Band  &  UVIT       & GALEX      & UVIT Flux & GALEX Flux   & SFR (UVIT)$^a$     & Log(SFR)\\
            &         & Exposure    &  Exposure  & density   & Density      & M$_{\odot}yr^{-1}$ &         \\
            &         &   time (Ks) & time (Ks)  & (Jy)      & (Jy)         &                    &         \\
\hline
ESO406-042  & FUV     & 1.994    & 0.281   & 2.10E-03$\pm$4.87E-04  & 2.05E-03$\pm$1.89E-05  &  0.089$\pm$0.021 & -2.06 \\
NGC 2541    & FUV     & 1.992    & 0.106   & 1.55E-02$\pm$3.87E-03  & 1.09E-02$\pm$1.00E-04  &  0.317$\pm$0.056 & -0.50 \\
NGC 5832    & FUV     & 1.972    & 0.111   & 5.08E-03$\pm$2.04E-03  & 5.35E-03$\pm$2.46E-04  &  0.056$\pm$0.022  & -1.25 \\                
\hline
\end{tabular}
\tablenotes{(a)~The SFR(UVIT) is calculated using the FUV and NUV intensities from table~2 and using the formula from \citet{salim.etal.2007} and the filter information from Tandon et al. (2020). For NGC 2541, since the FUV filter width was quite different compared to the GALEX filter, we normalised the CPS to GALEX CPS using background stars (see text). However, it should be noted that the SFRs have been corrected only for Galactic (Milky Way) extinction and not for the dust extinction arising from within the individual galaxies. Hence the true SFRs could be slightly higher.}
\end{table*}

\section{Results}

{\bf (i)}~The UV emission and the stellar disks~:~The UV emission from galaxies traces hot, young star forming regions for at least 10 times longer timescales than H$\alpha$ emission \citep{bianchi.2011}. In general, UV traces star formation for $\sim$10$^8$years \citep{thilker.etal.2007}. A small but significant fraction also traces low mass, helium-core burning stars on the horizontal branch of the main sequence; this is often called the UV-upturn and is commonly seen in bright elliptical galaxies \citep{yi.etal.2011}. There is also some UV emission from post asymptotic giant branch stars \citep{montez.etal.2017} and white dwarfs \citep{sahu.etal.2019}. The FUV emission arises from mainly hot stars, of which $\sim$40\% is from stars younger than 10~Myr and $\sim$60\% from stars that are older than 10~Myr \citep{calzetti.2013}. The low sky background in FUV and low number of foreground stars that are bright in FUV ensures a high stellar contrast. This makes FUV emission an ideal wavelength to detect small star forming regions that are not detected in H$\alpha$, especially since O-type stars account for $\sim$90\% of the FUV emission \citep{bianchi.etal.2014}. However, FUV sources account for only $\sim$10\% of the UV sources in GALEX surveys. Most of the UV emission from galaxies is composed of NUV emission from the redder population of stars (older and cooler). Hence, the NUV emission from galaxies is always more smoother and extended compared to the FUV emission. This is true for the 3 galaxies in our observations as well (Figures 1, 2, 3), especially for ESO406-042 and NGC 5832. However, for NGC2541 the FUV emission appears brighter. This maybe because of a shorter exposure time, but we note that both FUV and NUV exposures were of similar time (Table~2), and the filters are different and hence have different wavelength coverage \citep{rahna.etal.2017}. Some of the NUV data could be more noisy. Or there could be more young star formation in this galaxy compared to the other two galaxies. The fluxes are listed in Table~2. 
\\
{\bf (ii)}~The nature of the underlying stellar disks~:~When we compare the UV emission with the near-infrared (NIR) emission, we find that the stellar disks are surprisingly diffuse. This is because the stellar disks in the XUV regions have low surface densities, or in other words the number density of stars in these regions is low. Even in regions supporting many compact star forming knots, the underlying stellar disk as traced by the NIR emission appears to be very low. This is typical of type~2 XUV disks as noted in earlier studies \citep{thilker.etal.2007}. It is also seen in dwarfs where there is often a very low surface density disk underlying an extended HI gas disks \citep{patra.etal.2016,das.etal.2019}. When the mass surface density is plotted against the disk radius, it is seen to rapidly fall near the optical radius (e.g. NGC 628, NGC 3184) \citep{das.etal.2020}. This has important implications for star formation in these regions as a lower stellar disk surface density means the disk has less self gravity and so cloud collapse maybe affected.   
\\    
{\bf (iii)}~Morphology of star formation~:~The UV morphology traces the star formation in galaxy disks. In all three  galaxies the star formation is distributed over the diffuse stellar disks and does not follow any spiral arms, except for the northern arms of NGC 2541 where the star formation lies along elongated structures resembling arms. The star formating complexes are generally small and compact in nature (e.g. NGC 5832) or localized into large knots that are distributed over the entire disk (e.g. ESO406-042, NGC 2541).
\\  
{\bf (iv)~}Comparing the UV and HI emission~:~The HI emission in all 3 galaxies is considerable and fairly extended (Figure 5 and 6). The HI appears to be more concentrated in the inner region around the UV emisison. However, surprisingly the HI is not as extended as observed in type~1 XUV galaxies (e.g. NGC 628). This suggests that the cold gas is denser in the centers of type~2 XUV galaxies compared to type~1 XUVs, although the stellar disks in the XUV regions are very faint in both type~1 and type~2 XUVs. The dense HI is able to support the star formation. The molecular hydrogen (H$_2$) is probably localized around the compact star forming regions and hence not easy to detect \citep{bicalho.etal.2019}. This is evident from previous detections of very compact regions of molecular gas in the XUV disk of M63 \citep{dessauges-zavadsky.etal.2014}. 

\section{Discussion} 
It is clear from the comparison of the FUV and NUV images with the 3.6$\mu$m images that the underlying stellar disk in these star forming regions is very diffuse i.e. low in surface density. This is surprising as the star formation is widespread in all the galaxies. It is also seen in smaller dwarf galaxies as well \citep{das.etal.2019}, where an outer blue disk extends well beyond an inner old stellar disk which supports most of the star formation. It is known that the main triggers for outer disk star formation in XUV galaxies are (i) galaxy interactions (e.g. NGC 4625), (ii) gas accretion from high velocity clouds (HVCs) (e.g. NGC 891) and (iii) cold gas accretion by galaxies from the cosmic web (eg. NGC 2403) \citep{deblok.etal.2014}. However, not all galaxies with extended HI disks have XUV disk star formation nor do they all show signs of gas accretion from HVCs or the inter-galactic medium (IGM). Hence, it is still unclear why some galaxies show extended disk star formation and some do not, despite being rich in HI gas.

One possibility is that disk dark matter plays an important role in both supporting the HI disk in regions of low stellar mass density as well as increasing the disk mass surface density so that the Toomre instability factor Q reduces and instabilities can form in the diffuse stellar disks. The onset of local instabilities leads to star formation \citep{das.etal.2020}. This can explain the widespread star formation in type~2 XUV disks and the compact nature of the star formation as well. The dark matter associated with the disk could be due to the presence of a very oblate halo (Das et al. in preparation) or it may have resulted from the accretion of dark matter from several episodes of minor mergers during the evolution of the galaxy. 







\section{Conclusion}
1. We present UVIT FUV and NUV images of 3 XUV type galaxies, ESO406-042, NGC 2541 and NGC 5832. The first two are type~2 XUV galaxies and the third one is a mixed type XUV galaxy.\\
2. We also present the  GMRT HI maps of NGC 2541, both the intensity map and the velocity field. We also compare the UV emission from NGC 5832 with the Westerbrock archival HI map. All three galaxies are rich in HI and the star formation is associated with the cold gas emission. However, the HI emission is not as extended as in type~1 XUV galaxies.\\
3. The star formation is distributed over the diffuse stellar disks of the galaxies and is comprised of compact knots of star forming complexes. This suggests that the star formation originated in local instabilities in the galaxy disks rather than global instabilities such as spiral arms or bars.


\vspace{1cm}
\noindent
{\bf Acknowledgements} 
The authors gratefully acknowledge the IUSSTF grant JC-014/2017, which enabled the authors MD, NNP, and KSD to visit CWRU and develop the science presented in this paper. This publication uses the data from the UVIT, which is part of the AstroSat mission of the Indian Space Research Organisation (ISRO), archived at the Indian Space Science Data Centre (ISSDC). We gratefully thank all the members of various teams for providing support to the project from the early stages of design to launch and observations in the orbit. The HI observations were done using the GMRT. We thank the staff of the GMRT that made these observations possible. The GMRT is run by the National Centre for Radio Astrophysics of the Tata Institute of Fundamental Research. This research has used Spitzer 3.6micron images. This research has also made use of the NASA/IPAC Extragalactic Database (NED), which is operated by the Jet Propulsion Laboratory, California Institute of Technology, under contract with the National Aeronautics and Space Administration.
\\
{\bf Facilities:} Astrosat(UVIT), GALEX, GMRT, WSRT, Spitzer, SDSS, GBT, Parkes.


\bibliographystyle{aasjournal}
\bibliography{mdas_jaa2020}
\vspace{-1.5em}



\end{document}